\documentclass[fleqn,10pt]{wlscirep}
\usepackage{braket}
\usepackage{soul}
\usepackage{tabu}
\title{Nonlocality in Remote State Preparation vis-\`{a}-vis Teleportation}

\author[1]{Aiman Khan\thanks{aim15uph@iitr.ac.in}}
\author[2]{Som Kanjilal\thanks{som.kanji@jcbose.ac.in}}
\author[2]{Dipankar Home\thanks{dhome@jcbose.ac.in}}

\affil[1]{Department of Physics, Indian Institute of Technology-Roorkee, Roorkee - 247667, India.}

\affil[2]{Center for Astroparticle Physics and Space Science (CAPSS), Bose Institute, Kolkata - 700091, India.}


\begin{abstract}
The hitherto unexplored nonlocality of quantum correlations in the information transfer protocol of remote state preparation (RSP) is investigated in terms of a unified nonlocality argument formulated for teleportation as well as for two distinct RSP schemes. The argument differs only in the respective types of local measurements used in each scenario. One of the RSP schemes uses single particle projective measurements, while the other uses joint Bell-basis measurements. Interestingly, the quantum mechanical violations of the Bell-CHSH type inequalities formulated for each of the RSP setups, while being equal to each other, also turn out to be equal to that of the corresponding Bell-CHSH type local realist inequality for teleportation, using the common resource of non-maximally entangled state. This reveals that for such a resource state, the maximum amount of extractable nonlocality of correlations involved in the transfer of information is the same for the standard teleportation protocol and the two RSP schemes considered here, thereby implying that the essential difference between teleportation and remote state preparation lies in the amount of classical information required to be exchanged between parties, and not in the nonlocality of underlying correlations. 
\end{abstract}
\begin{document}

\flushbottom
\maketitle
%
%
\thispagestyle{empty}

\section*{Introduction}

Entanglement is an outstanding feature of quantum mechanics\cite{1,schrodinger1935discussion,bohmbook}. It lies at the heart of quantum information transfer protocols such as dense coding\cite{bennett1992communication} and teleportation\cite{bennett1993teleporting}, where it gives rise to an enhanced capacity of the communication channel between two spatially separated parties, beyond the classically attainable level. On the other hand, a profound implication of entanglement is quantum nonlocality, which is the incompatibility between an appropriately derived local realist constraint and the quantum mechanical correlations between the measured local observables pertaining to the spatially separated, entangled parties. This is usually shown in an archetypal setup where the quantum mechanically calculated correlations between the outcomes of local projective measurements made on each side violate the upper bound of the Bell-CHSH inequality\cite{bell1964instein,clauser1969proposed,bell1971introduction,clauser1974experimental}, or that for any of the other local realist inequalities pertaining to such a setup\cite{roy1978experimental,roy1979completeness,garuccio1980systematic,collins2004relevant}. In recent years, for revealing aspects of quantum nonlocality, different types of setups or manipulations on the two sides of the entangled system have been proposed\cite{popescu1995bell,gisin1996hidden,verstraete2002entanglement,hirsch2013genuine,vertesi2010two,barra2012higher}. These studies underscore the point that the manifestation of nonlocality is dependent on the measurement scenario as much as it is on the state itself.

Against the backdrop of the above line of investigations, the extent to which nonlocality can be revealed in the context of the setups used for entanglement-based information transfer protocols is, thus, a question of considerable significance. Here we note that the nonlocality of quantum teleportation, which involves the transmission of an arbitrary quantum state over space-like separation using local measurements and two bits of classical information, has been probed to some extent by a number of authors \cite{popescu1994bell,zeilinger1997quantum,hardy1999disentangling,zukowski2000bell,clifton2001nonlocality,barrett2001implications}. However, surprisingly, the nonlocality involved in any other information transfer protocol has remained unexplored. In this paper, we concentrate on the nonlocality of quantum correlations used for remote state preparation(RSP), which is considered in the following specific sense. 

Suppose Alice and Bob initially share a non-maximally entangled state $\cos\theta\ket{00} + \sin\theta\ket{11}$, characterized by the real variable $\theta$ which is accessible to Alice and Bob through their respective local measurements. Here Alice's task is to create a pure state at Bob's end, on the great circle characterised by the value of polar angle equal to $2\theta$ on Bob's Bloch sphere. The state itself is specified by the value of the additional parameter $\phi$. In order to achieve this, Alice uses local operations and classical communication which enable her to transfer information about the azimuthal angle $\phi$ on the aforementioned great circle, leading to the creation of the desired state $\cos\theta\ket{0} + \sin\theta e^{i\phi}\ket{1}$ at Bob's end. 

For realising the above kind of constrained remote preparation of states, we consider two distinct deterministic schemes which are are outlined in the following section. Subsequently, we analyse their nonlocality within a general framework which is first illustrated with respect to teleportation using non-maximally entangled states, and subsequently for the RSP schemes. This is followed by a comparison of the nonlocality of the RSP schemes with that of teleportation that enables a deeper understanding of the nature of relationship between them. The paper is concluded by summarising the insights gained into the nonlocality of both teleportation and the RSP schemes considered here, and directions for future studies are indicated.

\section*{Remote State Preparation of Constrained States}

The two specific RSP schemes discussed in this section can be used to prepare the same set of states at Bob's end, but by using different local operations at Alice's end. We first consider a scheme adapted from Lo's paper\cite{lo2000classical} that involves local von-Neumann projective measurements at Alice's end. Then, a novel RSP scenario is formulated where Alice makes joint Bell-basis measurement using an ancilla on her side to perform the same task. The former scheme is first discussed, as follows.

\subsection*{Remote State Preparation with von-Neumann Projective Measurement}

Suppose Alice and Bob share the following non-maximally entangled state:

\begin{equation}
    \ket{\Psi} = \cos\theta\ket{00} + \sin\theta\ket{11} 
\end{equation}

As remarked before, both Alice and Bob can individually determine the value of the parameter $\theta$. Alice's task is to prepare a pure state at Bob's end involving the phase factor $\phi$. In order to achieve this, Alice first makes the local rotation $\ket{1}\rightarrow e^{i\phi}\ket{1}$ on her particle after which the state becomes

\begin{equation}
    \ket{\Psi_{I}} = \cos\theta\ket{00} + e^{i\phi}\sin\theta\ket{11} 
\end{equation}

Alice then does a Hadamard transformation that corresponds to the following rotation:

\begin{equation}
\ket{0} \rightarrow \frac{1}{\sqrt{2}}(\ket{0} + \ket{1}), \hspace{5mm} \ket{1} \rightarrow \frac{1}{\sqrt{2}}(\ket{0} - \ket{1})
\end{equation}

which converts the entangled state given by $\Psi_{I}$ of Eq.2 to 

\begin{equation}
\ket{\Psi_{II}} = \frac{1}{\sqrt{2}} \ket{0} (\cos\theta\ket{0} + \sin\theta e^{i\phi}\ket{1}) +  \frac{1}{\sqrt{2}} \ket{1}(\cos\theta\ket{0} - \sin\theta e^{i\phi}\ket{1})
\end{equation}

If Alice now makes a von-Neumann projective measurement in the $\sigma_{z}$ computational basis, it is easily seen that Bob's state is either $\cos\theta\ket{0} + e^{i\phi}\sin\theta\ket{1}$ or $\cos\theta\ket{0} - e^{i\phi}\sin\theta\ket{1}$, depending on whether Alice gets the outcomes +1 or -1 (which are eigenvalues of $\sigma_{z}$ corresponding to $\ket{0}$ and $\ket{1}$ respectively). 

Alice then sends the results of her measurements to Bob, encoded onto one bit (corresponding to the two outcomes she obtains). Based on this communication, Bob can sort his states into the two subensembles of pure states mentioned above. We also see that since these two states can be converted into each other by $\ket{1} \rightarrow -\ket{1}$, which is a simple phase flip, Bob can always always ends up with the state that Alice wishes to create, thereby ensuring the scheme to be deterministic.    

\subsection*{Remote State Preparation with Bell-basis Measurement}

We now propose a scheme to remotely prepare the same final state as in the previous scheme, using the same entangled state given by Eq.1, but based on joint measurements at Alice's end, instead of single-particle von-Neumann projective measurement. For this purpose, Alice first performs the local phase transformation, $\ket{1}\rightarrow e^{i\phi}\ket{1}$, which fixes the parameter $\phi$ corresponding to the state on the great circle specified by polar angle value of $2\theta$ that will be created at Bob's end. Subsequently, Alice applies the C-NOT gate on her qubit and an ancilla which is introduced at her end, with Alice's qubit as the control bit. Consequently, Eq.1 becomes 

\begin{equation}
\ket{\Phi_{I}} = U_{\mbox{An-A}}\ket{\xi}\otimes\ket{\Psi} = \cos\theta\ket{0}\ket{00} + \sin\theta e^{i\phi}\ket{1}\ket{11}
\end{equation}
where $\ket{\xi}$ is the state of the ancilla. The scheme does not depend on the choice of the ancilla state, as we will show soon. For brevity, here the state of the ancilla is taken to be one of the computational basis states ($\ket{0}$). $U_{\mbox{An-A}}$ is the unitary C-NOT operation defined by the following transformation(with the second qubit as the control bit):

\begin{equation}
    \ket{x}\ket{0} \rightarrow \ket{x}\ket{0}, \hspace{5mm}
    \ket{x}\ket{1} \rightarrow \ket{(x+1)\,\mbox{mod}\,2}\ket{1}, \hspace{3mm} x\in\{0,1\}
\end{equation}

Alice then does Bell-basis measurements on her qubit and the ancilla, and sends information to Bob about the outcomes. If we express the kets in the ancilla-Alice space in terms of the Bell-basis states, Eq.5 can be written as :

\begin{equation}
\frac{1}{\sqrt{2}} \ket{\phi^+} (\cos\theta\ket{0} + \sin\theta e^{i\phi}\ket{1}) + \frac{1}{\sqrt{2}} \ket{\phi^-} (\cos\theta\ket{0} - \sin\theta e^{i\phi}\ket{1})
\end{equation}
where 
\begin{equation}
    \ket{\phi^{\pm}} = \frac{1}{\sqrt{2}}(\ket{00} \pm \ket{11}), \hspace{5mm} \ket{\psi^{\pm}} = \frac{1}{\sqrt{2}}(\ket{01} \pm
    \ket{10})
\end{equation}

From the form of Eq.7, it is seen that Alice will only get two measurement outcomes corresponding to the Bell-states $\{\phi^+,\phi^-\}$, which can be encoded onto a single bit and communicated to Bob. Based on this communication, Bob can then separate his ensemble of states into two subensembles of pure one-qubit states that are inter-convertible in the same way as discussed as the last subsection, thereby ensuring that this scheme is also deterministic.

Let us now consider the above scheme by starting  with a general ancilla state $\ket{\xi} = a\ket{0} + b\ket{1}$ and going through the same steps as above. Alice will then get outcomes corresponding to all the four Bell-basis states. The counterpart of Eq.7 will be

\begin{eqnarray}
    \frac{a}{\sqrt{2}} \ket{\phi^+} (\cos\theta\ket{0} + \sin\theta e^{i\phi}\ket{1}) \nonumber &+& \frac{a}{\sqrt{2}} \ket{\phi^-} (\cos\theta\ket{0} - \sin\theta e^{i\phi}\ket{1}) + \nonumber \\
    \frac{b}{\sqrt{2}} \ket{\psi^+} (\cos\theta\ket{0} + \sin\theta e^{i\phi}\ket{1})  &+& \frac{b}{\sqrt{2}} \ket{\psi^-} (\cos\theta\ket{0} - \sin\theta e^{i\phi}\ket{1}
\end{eqnarray}

Eq.9 shows that Bob's conditional state is the same for both projectors $\{\ket{\phi^+}\bra{\phi^+}$ and $\ket{\psi^+}\bra{\psi+}\}$, and for $\{\ket{\phi^-}\bra{\phi^-}$ and $\ket{\psi^-}\bra{\psi-}\}$. This allows Alice to club her outcomes into two pairs, effectively making it a two-outcome measurement. The probability of each of these two outcomes can be shown to be $\frac{1}{2}$, independent of the parameters of the ancilla state. This clearly shows that, whatever be the ancilla state, this protocol works equally well and is deterministic.

Here we would like to remark that the above scheme is teleportation-like, in the sense that Bell-basis measurements(on Alice and an ancilla) and classical communication are used. This feature will be important for later discussions comparing the nonlocality in RSP with that in teleportation.

\section*{A Unified Non Locality Argument for Teleportation and Remote State Preparation}

A typical information transfer protocol involves local operations/measurements and transfer of classical information between spatially separated parties that can communicate with each other. Here it is important to note that the role of classical communication is to enable the parties to access the transferred information, while the transfer of information is accomplished irrespective of the classical communication bits\cite{zeilinger1997quantum}. The question as to whether the quantum correlations that enable this information transfer are incompatible with any local realist model is what we mean by the nonlocality of such a protocol. For example, in teleportation, the nonlocality of correlations used for transfer of information is analysed by getting rid of the classical channel and considering the following: Alice, the sender, still makes Bell-basis measurements but can vary her choice of input ancilla state, while the receiver, Bob, can do spin projective measurements along the direction he chooses. Let us now delve deeper into the analysis of the nonlocality of teleportation in the next section, which will set the stage for  discussion of nonlocality in the context of remote state preparation.

\subsection*{Nonlocality of Teleportation using Non-Maximally Entangled States}

To recall for completeness, teleportation is the transfer of an arbitrary ancilla state, say $\ket{\eta}$, using an entangled resource state we will call D shared between Alice and Bob. Alice performs joint Bell-basis measurement on her particle and the ancilla, and sends the outcomes of her measurement to Bob via two bits of information. Based on this information, Bob performs unitary rotations on his particle according to some pre-agreed strategy. The original teleportation scheme used maximally entangled states for the resource state D\cite{bennett1993teleporting}, resulting in the ancilla state being perfectly transferred to Bob every single time. However, we can continue to use the same protocol (which includes the strategy used to assign Bob's unitary rotations to Alice's outcomes) when the resource state is of the non-maximally entangled form Eq.1, in which case the ancilla state is reconstructed imperfectly. If the state produced at Bob's end is $\rho_{\mbox{Bob}}$, we define fidelity of the transfer F as $\mbox{M}\braket{\eta|\rho_{\mbox{Bob}}|\eta}$, where M stands for mean over all the possible ancilla states $\ket{\eta}$. This has been shown to be\cite{gisin1996nonlocality}:
\begin{equation}
F(\theta) = \frac{2}{3}(\cos^3\theta - \sin^3\theta)/(\cos\theta-\sin\theta)
\end{equation}
where we see that F($\theta=\pi/4)=1$ and it goes to the classical value of $2/3$ as we move towards product states\cite{popescu1994bell}.

Let us now analyse the nonlocality of such an information transfer. We suppose Alice does joint Bell-basis measurements for the two choices of ancilla states she wishes to teleport, $\ket{\eta_1}$ and $\ket{\eta_2}$, while Bob measures spin along two different directions, $\hat{n}_1$ and $\hat{n}_2$. We further introduce the following bivalent functions of Bell-basis projectors on ancilla-Alice subspace:

\begin{eqnarray}
  A_1 &=& \hspace{3mm} \ket{\phi_-}\bra{\phi_-} + \ket{\psi_-}\bra{\psi_-} - \ket{\phi_+}\bra{\phi_+} - \ket{\psi_+}\bra{\psi_+}\nonumber \\
  A_2 &=&  -\ket{\phi_-}\bra{\phi_-} + \ket{\psi_-}\bra{\psi_-} + \ket{\phi_+}\bra{\phi_+} - \ket{\psi_+}\bra{\psi_+}\nonumber,
\end{eqnarray}
whose values (which are $\pm1$) Alice can determine from the outcome of her Bell-basis measurement, as can be seen easily. Like the Bell-CHSH scenario, we now choose two settings on each side -- while the choices on Bob's side have been already stated, Alice chooses to determine either the value of the operator $\mbox{A}_1$ while she teleports $\ket{\eta_1}$, or the value of $\mbox{A}_2$ while teleporting $\ket{\eta_2}$. The usual derivation of the Bell-CHSH inequality\cite{bell1971introduction,clauser1974experimental}, adapted to this scenario, then gives us the following expression:

\begin{equation}
B_1 =  |\langle A_1\otimes\sigma_1\rangle_{\ket{\eta_1}\bra{\eta_1}\otimes D} \,\,\,+\,\,\, \langle A_1\otimes\sigma_2\rangle_{\ket{\eta_1}\bra{\eta_1}\otimes D} \,\,\,+\,\,\, \langle A_2\otimes\sigma_1\rangle_{\ket{\eta_2}\bra{\eta_2}\otimes D} \,\,\,-\,\,\,  \langle A_2\otimes\sigma_2\rangle_{\ket{\eta_2}\bra{\eta_2}\otimes D}| \,\,\,\leq\,\,\, 2 
\end{equation}
where $\sigma_1 = \mathbf{\sigma}.\hat{n}_1$, $\sigma_2 = \mathbf{\sigma}.\hat{n}_2$, and D is the entangled state of Alice and Bob. 

A violation of this inequality implies the nonexistence of a local hidden-variable model where Bob's measurement outcome is independent of the choice of state Alice attempts to teleport. The local realism used in obtaining Eq.11 also assumes that Alice's measurement outcomes are independent of Bob's choice of the direction of projective measurement. Thus, violation of these local realist notions in this scenario is what is entailed by the nonlocality of correlations used in teleporting the state. These implications,we stress, are distinct from the non-locality of correlations in the standard Bell test where a violation implies the impossibility of an LHV model explaining correlations between arbitrary measurements on both sides. A violation of the teleportation inequality in Eq.11 only implies the non-existence of a local model that mimics the correlations that actually participate in teleportation at the hidden variable level.

The LHS of the inequality in Eq.11 (denoted by $\mbox{B}_1$) depends on Alice's choice of ancilla states, $\ket{\eta_1}$ and $\ket{\eta_2}$, and on Bob's choice of the direction of projective measurements, $\hat{n_1}$ and $\hat{n_2}$, besides the resource state of Eq.1 characterised by $\theta$. We now plot the supremum of this function over measurement choices on each side, denoted thence by $B_1(\theta)$ in Fig 1(a). A violation of the local realist bound is seen only for $\theta\gtrapprox\pi/8$ (where the approximate inequality indicates that the value is calculated numerically). Teleportation using resource states characterised by $ \theta \gtrapprox \pi/8$ cannot therefore occur without involving nonlocality. As a corollary, for values of $\theta$ less than $\pi/8$, teleportation may not be non-local, although the underlying non-maximally entangled resource states violate the Bell-CHSH inequalities by themselves and hence are non-local. This feature has not yet been pointed out for teleportation using non-maximally entangled states. The corresponding fidelity beyond which teleportation occurs non-locally is $\frac{2}{3}(1 + 1/2\sqrt{2}) \simeq 0.90$, as follows from Eq.10, by putting $\theta=\pi/8$ in the $\mbox{F}(\theta)$ function. It is curious to note that this is also true if Werner states are used as resource, and correlations used are non-local only for values of fidelity greater than $\frac{2}{3}(1 + 1/2\sqrt{2})$\cite{clifton2001nonlocality}. 

Note that Eq.11 is only a particular form of a wider class of local realist inequalities that can be formulated in this scenario. Different forms of these inequalities depend upon the choice of joint observables (involving different linear combinations of Bell basis projectors, or even more general forms) measured by Alice. The specific form of Eq.11 appears in ref.24, where it was derived as a special case of the teleportation inequality appearing in ref.23.

It may be also be emphasised that the setup indicated above is an example of the generalised measurements one can make on the two parts of the entangled system to reveal the nonlocality present in a given scenario. Hence, teleportation itself, without the classical channel, can be looked upon as a non-trivially different setup for showing quantum nonlocality where the underlying nonlocality is that of the state contingent upon the scenario in which it is manifested.

Given the above background, we now proceed to address the issue of nonlocality for each of the two schemes for remote state preparation discussed in the previous section. A key point in discussing nonlocality in teleportation was the local realist assumption, among others, of the independence of the outcome of Bob's measurement with respect to the choice of the state Alice chooses to teleport to Bob. A bound on the amount of local correlations in both the RSP schemes considered in this paper follows similarly from an assumption of independence of Bob's measurement outcome with respect to the choice of Alice's local phase transformation, specified by the variable $\phi$, and consequently of the state (on the great circle corresponding to a given polar angle $2\theta$) that Alice wishes to create at Bob's end. 

\subsection*{Nonlocality of Remote State Preparation with von Neumann Projective Measurement}

\begin{figure}[ht]
\centering
\includegraphics[scale=0.042]{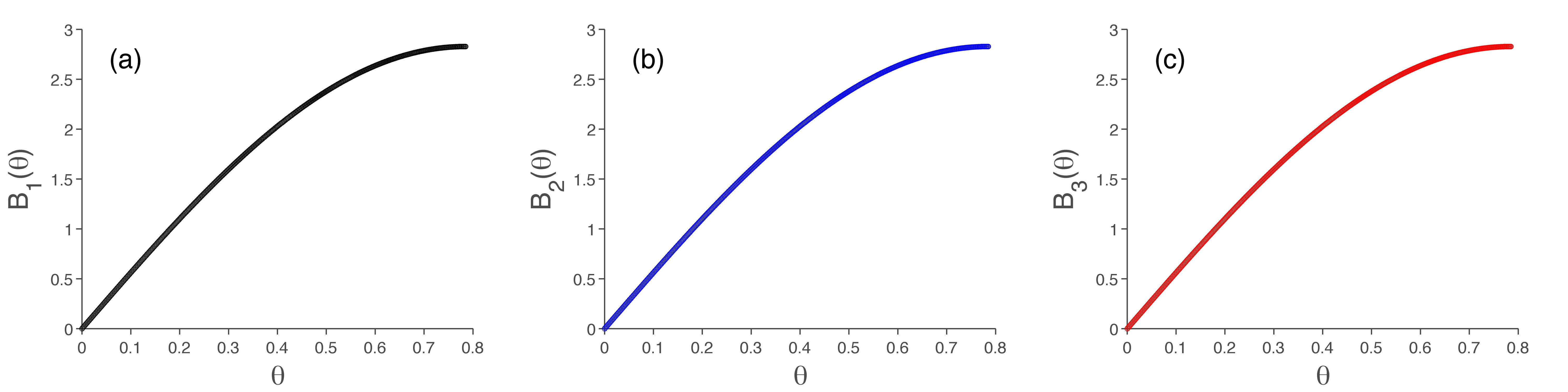}
\caption{The correlators appearing in Eqs. (11), (12) and (13), corresponding to teleportation and the two RSP schemes presented respectively, maximized over all possible measurements settings (as defined separately for each scheme), are plotted in (a), (b) and (c) (in that order) against the parameter $\theta$ defining the initial entanglement of Alice and Bob's joint state.}
\end{figure}

In order to formulate the nonlocality argument of the RSP protocol using von Neumann projective measurements (discussed in the first subsection of the previous section), analogous to teleportation, we set up the appropriate scenario by first getting rid of the classical communication. In contrast to the usual Bell scenario where Alice can choose the direction of her spin projective measurement, here she chooses the parameter of her local phase transformation while continuing to perform the unitary Hadamard transformation on her particle and the subsequent measurement of the $\sigma_z$ observable. In the absence of classical communication, Bob does a spin projective measurement along a direction of his choice. Alice and Bob can then combine the outcomes of their measurements to test the nonlocality of the correlations between the outcomes. An important element in our nonlocality argument is that, in RSP, Alice has the freedom to choose the azimuthal angle $\phi$ of the state to be prepared at Bob's end, which is like the freedom to choose the state that Alice can teleport in teleportation. 

We can now develop a the relevant local realist inequality based on this scenario analogous to the case of teleportation (described in detail in the last subsection). The measurement setting on Alice's side is defined by her choice of the parameter $\phi$, while Bob chooses the direction of his spin projective measurement. With two distinct settings thus specified on each side, the following Bell-CHSH type inequality will then hold:

\begin{equation}
\mbox{B}_2(\theta,\phi_1,\phi_2,\hat{n}_1,\hat{n}_2)\,\, = \,\,|\langle \sigma_z\otimes\sigma_1\rangle_{\ket{\Psi(\phi_1)}\bra{\Psi(\phi_1)}} \,\,+\,\, \langle \sigma_z\otimes\sigma_2\rangle_{\ket{\Psi(\phi_1)}\bra{\Psi(\phi_1)}} \,\,+\,\, \langle \sigma_z\otimes\sigma_1\rangle_{\ket{\Psi(\phi_2)}\bra{\Psi(\phi_2)}} \,\,-\,\, \langle \sigma_z\otimes\sigma_2\rangle_{\ket{\Psi(\phi_2)}\bra{\Psi(\phi_2)}}|  \,\,\leq\,\, 2 
\end{equation}
where $\sigma_1 = \mathbf{\sigma}.\hat{n}_1$, $\sigma_2 = \mathbf{\sigma}.\hat{n}_2$ and $\theta$ is the variable characterising the degree of entanglement of the starting joint state of Eq.1 of the protocol. The expectation values are calculated with respect to the joint  wave function of Alice and Bob after the Hadamard transform, given by Eq.4, denoted here by $\ket{\Psi(\phi)}$ to emphasise the $\phi$-dependence.

We plot the value of the Bell operator in Eq.12, optimized over all possible choices of $\phi$(Alice's end) and $\hat{n}_1$,$\hat{n}_2$(Bob's end), as a function of $\theta$ (denoted by $\mbox{B}_2(\theta)$) in Fig 1(b). As $\theta$ is varied from $0$ to $\pi/4$, Alice and Bob's initial resource state goes from a shared product state to the maximally entangled state. The plot generated in Fig 1(b) is the same as the one we obtained for teleportation given by Fig 1(a), with a violation of the local realist bound for $\theta \gtrapprox \pi/8$ (where the approximate inequality similarly indicates that the value is calculated numerically), and a maximum violation of $2\sqrt{2}$ for $\theta=\pi/4$, corresponding to maximally entangled resource state. 

\subsection*{Nonlocality of Remote State Preparation with Bell-basis Measurement}

We now examine the nonlocality of the scheme for RSP with Bell-basis measurements (introduced in the second subsection of the last section). The scheme uses joint Bell-basis measurements on Alice's end to effect the disentanglement required to prepare states remotely at Bob's end, as opposed to the single-particle projective measurements in the previous scheme. As pointed out before, this protocol is teleportation-like. Therefore, to study its nonlocality, we can use the Bell-CHSH type teleportation inequality of Eq.11, modified appropriately to fit the following scenario : Alice can choose her phase rotation, which boils down to her choice of the parameter $\phi$, while continuing to perform the subsequent C-NOT operation and Bell-basis projective measurement, as per the protocol. Bob, in the absence of the classical communication, makes a single-particle ideal spin measurement on his particle along the direction of his choice. In the same vein as that for the two preceding schemes, the correlations that enable the information transfer would have to satisfy the following local realist inequality, if describable by a local realist model of this scenario:

\begin{equation}
\mbox{B}_3(\theta,\phi_1,\phi_2,\hat{n}_1,\hat{n}_2) = |\langle A_1\otimes\sigma_1\rangle_{\ket{\Psi(\phi_1)}\bra{\Psi(\phi_1)}} \,\,\,+\,\,\, \langle A_1\otimes\sigma_2\rangle_{\ket{\Psi(\phi_1)}\bra{\Psi(\phi_1)}} \,\,\,+\,\,\, \langle A_2\otimes\sigma_1\rangle_{\ket{\Psi(\phi_2)}\bra{\Psi(\phi_2)}} \,\,\,-\,\,\,  \langle A_2\otimes\sigma_2\rangle_{\ket{\Psi(\phi_2)}\bra{\Psi(\phi_2)}}| \,\,\,\leq\,\,\, 2 
\end{equation}
where the terms carry the same meaning as the previous subsection, and the expectation values are calculated with respect to the grand wave function of the ancilla-Alice-Bob system after the C-NOT operation, given by Eq.5, denoted here by $\ket{\Psi(\phi)}$ to emphasise the $\phi$-dependence.

Again, we plot the value of the Bell correlator function given by the LHS of the above inequality as a function of $\theta$ (denoted by $\mbox{B}_3(\theta)$) in Fig. 1(c), maximizing over all possible measurement settings on the both sides, just like in the two previous subsections. We showed in the previous section that starting with a given state, either RSP scheme can be used to prepare the same final states with the same one bit of classical communication. Given this equivalence, unsurprisingly, we find that the plot for $\mbox{B}_3(\theta)$ overlaps exactly with that for $\mbox{B}_2(\theta)$ and therefore with $\mbox{B}_1(\theta)$. 

\subsection*{Comparison of these Non-Localities}

At the outset, we must first clarify that the specific forms of the local realist inequalities appearing in Eqs. 11, 12 and 13 are obtained in the context of a common Bell-CHSH type scenario and are distinct only in the details of the measurements made on either side, which is why a meaningful comparison of the nonlocality of these schemes in terms of these inequalities is possible. The central result that emerges from this comparison is that the maximum value of the Bell correlators appearing in the inequalities is the same for teleportation and the two RSP schemes, if we use any non-maximally entangled resource state. This is reflected in the fact that the three curves for $\mbox{B}_1(\theta)$, $\mbox{B}_2(\theta)$ and $\mbox{B}_3(\theta)$ in Fig.1 are the same, and implies the sameness of the amount of nonlocality which can be extracted in each of these schemes.  

In teleportation, the decreasing nonlocality of correlations (measured in terms of the value of $\mbox{B}_1(\theta)$), as $\theta$ decreases, can be understood in terms of its trade-off with fidelity -- lesser amount of nonlocality would mean a greater probability of error in teleporting a given state and a decreasing fidelity. On the other hand, the RSP protocol considered in this paper is deterministic for all values of $\theta$ and hence the amount of nonlocality involved seems to have no bearing on the probability with which the task is successfully accomplished. However, nonlocality seems to play a subtle role in RSP suggested by the feature that the cardinality of the set of states on the great circle characterised by polar angle $2\theta$ increases as we go from the poles to the equator and, hence, more information (whether quantum or classical) would be needed to specify them. Now, since the classical information transferred is fixed to be one bit, more information transfer would be required in the quantum channel, which would suggest that a greater amount of nonlocality of correlations would be needed to prepare Bob's state on the equator than on the poles which, in turn, is reflected by the increasing values of $\mbox{B}_2(\theta)$ and $\mbox{B}_3(\theta)$ as $\theta$ increases.

Furthermore, from the same magnitude of the quantum violations of local realism in both teleportation and RSP, one is led to conclude that, for resource states of Eq.1, the essential difference between these two lies in the amount of classical information transferred in the respective final stages. It is interesting to note that for the special case of maximally entangled resource state, perfect teleportation faithfully relays information about the two parameters that specify the state of the ancilla to Bob, while in RSP, information about only one parameter (the azimuthal angle on the equator of the Bloch sphere corresponding to the state Alice is trying to prepare) is transferred to Bob. Thus, one would have intuitively expected that the  nonlocality of correlations required for perfect teleportation would be greater than that needed for remote preparation of a state on the equator of the Bloch sphere, which is not the case. Another curiosity to be noted is that the RSP information task can be carried out with perfect accuracy using either protocol presented without actually utilising non-locality, for resource states of Eq.11 with $\theta<\pi/8$. This is in stark contrast to teleportation where non-locality is essential to achieve faithful transfer.

\section*{Conclusions}

In a nutshell, this paper provides an overarching logical framework for analysing the role of nonlocality in implementing information transfer tasks in the context of various protocols, whose application is illustrated here specifically with respect to teleportation and remote state preparation. This opens up a range of possibilities for probing the previously unexplored nonlocality of correlations involved in other information transfer schemes such as in probabilistic teleportation\cite{agrawal2002probabilistic}, entanglement swapping\cite{bennett1993teleporting,zukowski1993event} and other schemes for remote state preparation\cite{pati2000minimum}. Such studies would not only enable comparing and contrasting the various information transfer protocols, but may also provide heuristic clues for harnessing nonlocality as a resource for more efficient transfer of quantum information. Here we may note that some studies have been done probing the use of nonlocality as a resource for information processing \cite{barrett2005nonlocal}, and the amount of resources required to simulate correlations in entangled quantum states\cite{popescu1994quantum,cerf2004quantum,toner2003communication,brunner2005entanglement}. 

It needs to be stressed that the nonlocality demonstrated in this paper is that of the correlations involved in the measurement scenario we impose on the quantum state pertinent to the information transfer protocol considered, which goes beyond the usual
Bell scenario involving projective single-particle measurements on either side of the apparatus. Substituting the single ideal measurements used in the usual Bell-CHSH scenario with local transformations and subsequent measurements required for the specific information transfer protocols of teleportation and RSP, a unified nonlocality argument is formulated in this paper using the common resource of non-maximally entangled states. Thus, the analysis provided in this paper may serve as a springboard for a comprehensive study of the role of nonlocality of quantum correlations as a resource actually involved in the information transfer in various protocols which, in turn, can enrich the earlier mentioned studies concerning nonlocality as a resource which only consider correlations between the outcomes of single particle projective measurements on the two sides of the entangled system. 

We may also emphasise that the simple, first principle approach we take to a Bell-type nonlocality argument for correlations involved in the information transfer can be adapted with similar ease to build non-CHSH type inequalities. For example, an I3322-type local realist inequality bounding correlations involved in the teleportation protocol may have the following form (considering Alice calculates $\mbox{A}_1$ when she is attempting to teleport ancilla states $\ket{\eta_1}$ or $\ket{\eta_2}$, and $\mbox{A}_2$ when she has $\ket{\eta_3}$ to teleport; other choices also possible):

\begin{eqnarray}
|\langle A_1\otimes\sigma_1\rangle_{\ket{\eta_1}\bra{\eta_1}\otimes D} &+& \langle A_1\otimes\sigma_1\rangle_{\ket{\eta_2}\bra{\eta_1}\otimes D} + \langle A_2\otimes\sigma_1\rangle_{\ket{\eta_3}\bra{\eta_1}\otimes D} + \langle A_1\otimes\sigma_2\rangle_{\ket{\eta_1}\bra{\eta_1}\otimes D} + \langle A_1\otimes\sigma_2\rangle_{\ket{\eta_2}\bra{\eta_1}\otimes D} \nonumber \\
&-& \langle A_2\otimes\sigma_2\rangle_{\ket{\eta_3}\bra{\eta_1}\otimes D} + \langle A_1\otimes\sigma_3\rangle_{\ket{\eta_1}\bra{\eta_1}\otimes D} -  \langle A_1\otimes\sigma_3\rangle_{\ket{\eta_2}\bra{\eta_1}\otimes D} - \langle A_1\rangle_{\mbox{tr}_{B}(\ket{\eta_1}\bra{\eta_1}\otimes D)} \nonumber \\
&-& 2\langle\sigma_1\rangle_{ \mbox{tr}_{A}(\ket{\eta_1}\bra{\eta_1}\otimes D)} - \langle\sigma_2\rangle_{ \mbox{tr}_{A}(\ket{\eta_1}\bra{\eta_1}\otimes D)}| \,\,\leq\,\, 0
\end{eqnarray}
where $A_1\otimes\sigma_1\rangle_{\ket{\eta_1}\bra{\eta_1}\otimes D}$ is the joint probability of Alice obtaining the result -1 for the measurement operator $A_1$ and Bob obtaining the outcome -1 for the operator $\sigma_1$, given that the shared state is ${\ket{\eta_1}\bra{\eta_1}\otimes D}$, and similarly for the other terms. Note that these are joint probabilities here and not expectation values as was the case before, and the notation has only been retained for brevity. It is quite remarkable that we can skip the complicated way in which Zukowski\cite{zukowski2000bell} derives the rather simple expression of the CHSH-type teleportation inequality, and derive from scratch not only the CHSH-type teleportation inequality but also the I3322-type teleportation inequality and similarly the RSP inequalities. Using inequalities of the type in Eq.14 for teleportation using the resource states D in Eq.1, we can check that the value of the I3322 correlators are the same for teleportation and the two RSP protocols, even though none of them violate local realism for any $\theta$. This result further underscores the sameness of nonlocality involved in teleportation and remote state preparation.

An upshot of our study is that the correlations involved in teleportation and RSP are both non-local for a range of non-maximally entangled states (for values of $\theta\gtrapprox\pi/8$) that is smaller than the set of states for which nonlocality is revealed in the usual Bell-CHSH scenario (any value of $\theta$). This result is not entirely unexpected in light of ref.24, where it was shown that if teleportation using Werner states is non-local, then the underlying resource state violates the CHSH inequality, but the converse is not true. It thus seems to be the case that it is easier to simulate the information theoretic tasks of teleportation and RSP using local realist models than to simulate the outcomes of single particle projective measurements in the usual Bell scenario, as was pointed out for teleportation using Werner states\cite{clifton2001nonlocality}.  

It is important to stress that the common procedure invoked for setting up the Bell-CHSH type inequality for teleportation and RSP lets us make a legitimate comparison of the nonlocality of correlations involved in the two processes\cite{forster2009distilling}. Just as the maximally entangled state is thought to be more non-local than a non-maximally entangled state because of its stronger violation of the Bell-CHSH inequality, the value of the Bell-CHSH type correlator for each scheme we construct serves as a measure of nonlocality of the correlations used, which is crucially contingent upon the context. In this case, the context constitutes the details of the measurements made on Alice's side which, in turn, are dependent upon the specifics of the respective protocols. 

The results of the above comparison are striking -- we find that both teleportation and RSP use the \textit{same} amount of non-local correlations for their respective transfers of information, and the only difference between them, as was pointed out in the previous section, is the amount of classical information Alice needs to communicate to Bob. This is an important conclusion, as it lets us pinpoint clearly the extent to which nonlocality is a resource in each of these schemes. 

Finally, we would like to mention that the particular measurement scenario we impose on the spatially separated parties to show nonlocality in this paper can be extended by using either more number of settings in the correlators \cite{roy1979completeness,garuccio1980systematic,collins2004relevant}, or by using generalised measurements\cite{vertesi2010two,barra2012higher} on either side. This line of study would require using appropriate versions of local realist inequalities to identify the sets of non-local correlations in a way more general than just using the Bell-CHSH type inequalities, and would also enable showing any possible inequivalence between the various local realist inequalities\cite{roy1978experimental,roy1979completeness,garuccio1980systematic,collins2004relevant} in the context of the information transfer protocols. Thus, a comprehensive study for assessing relative efficacies of the various local realist inequalities in analysing the nonlocality of correlations in teleportation and RSP, as well as in other schemes of quantum information transfer, should be a fruitful line of research yielding deeper insights into the whole issue of quantum nonlocality and its use as a resource.

\bibliography{bibliographyRSP}

\section*{Acknowledgments}

A.K wishes to thank Dr. Rajdeep Chatterjee for co-supervision of this work as part of his master's thesis at Bose Institute. D.H. acknowledges support from the Dept. of Science and Technology, Govt. of India (Grant Reference Number SR/S2/LOP08/2013) and from Centre for Science, Kolkata. 

\section*{Author contributions statement}

Authors A.K., S.K and D.H. contributed equally in conception of the idea, calculations therewith and analysis of the results. All authors reviewed the manuscript. 

\section*{Additional information}

\textbf{Competing financial interests} : The authors declare no competing financial interests.

\end{document}